# Could the Coandă effect be called the Young effect? The understanding of fluid dynamics of a legendary polymath

T. López-Arias [a)]

Department of Physics, University of Trento

38123 Povo (Trento) - Italy

A paper of Thomas Young (1773-1829) on the behavior of streams of air makes a good starting point for discussing the not always fair, sometimes serendipitous, association of a physical phenomenon with its discoverer. In a didactic context, the introduction of historic anecdotes and the particular details of a scientific discovery represent an effective tool for establishing interdisciplinary connections that may help in the learning process and may unveil unexpected insights in the disposition of a scientist. We discuss a small part of a famous paper by Young on sound and light. We show that the proverbial intuition of this famous polymath, applied to complex fluid dynamic phenomena, may lead to the discussion of the Coandă effect, the physical origin of lift, and the behavior of streams of air, as well as weave an interesting interplay of several crucial names in the history of aerodynamics.

## I. INTRODUCTION

Wilbur (1867-1912) and Orville (1871-1948) Wright hold the primacy of being the first men to have reached the ancient human dream of flying. In the words of Orville describing their first celebrated flight performed December 17, 1903, at Kitty Hawk: "The first of these flights … was the first time in the history of the world that a machine carrying a man and driven by a motor had lifted itself from the ground in free flight".[1] This is a very accurate definition of what is meant by flying, but it does not mean that the Wrights were the first to attempt being lifted in the air by means of a "heavier than air machine", as the famous book "Progress on flying machines", written by the railway engineer Octave Chanute (1832-1910) in 1894, thoroughly demonstrates.[2] In the introduction Chanute states that he will treat "of Flying Machines proper – that is to say to treat of forms of apparatus heavier than the air which they displace; deriving their support from and progressing through the air, like the birds, by purely dynamical means". The importance of this book in the Wright's success is historically acknowledged.[3] Chanute and the Wrights kept a long friendship, as he always provided them support and technical advice. He was often at Kitty Hawk, one of the first eye witnesses of the Wright's attempts with their first gliders. These were actually inspired by Chanute's own earlier projects. Chanute's book, with its thorough, state-of-the-art compilation on flying machines, represented a fundamental reference for all the "dreamers" who were devoting their lives to the actual possibility of flying.

In the first pages of this seminal book, Chanute gives a short overview of the elementary technical ideas about the physical basis of flight. For instance, he provides the definition of lift and "drift" (drag) forces, explains their vector composition and discusses the dependence of the pressure field over a given surface on its angle of incidence (angle of attack) with respect to the flow, an open problem at the time. A brief discussion follows in which the mechanism of interaction between solids and fluid flows is described. At this point, Chanute describes a "mysterious" phenomenon referring that "certain shapes when exposed to currents of air under certain ill-understood circumstances, actually move toward that current instead of away from it".[2] He continues describing how a hollow sphere suspended close to a jet of air will be pulled towards it, as is demonstrated by experiments performed by Thomas Young (1773-1829) in 1800. Chanute does not provide the reference, but he is certainly quoting the "Outlines of Experiments and Inquiries Respecting Sound and Light,"[4] in which Young declares the intention to present to the Royal Society "a few observations on the subject of sound". In these forty five pages, Young directs attention to, among other subjects, air discharge from apertures, the observation of the nature of sound through clever smoke visualization techniques, musical sounds, and the vibration of rods and plates. All these subjects are related to sound and the behavior of air streams. The letter ends with a chapter devoted to the temperament of musical instruments. The most important part of this letter is chapter X which is titled "Of the analogy between light and sound". In it Young exploits his observations on air streams and sound propagation to support the wave nature of light.

Today, the "mysterious" phenomenon described by Chanute is known as Coandă effect,[5] and it is accurately described by Young in his letter. It regards the adhesion of jets of air to curved surfaces, and it is usually demonstrated with the simple use of a ping-pong ball and a household hairdryer. As we will see below, Young gives a proper physical explanation for this phenomenon, demonstrating his accurate insight and indicating many other important phenomena related with the behavior of jets of air.

Chanute describes the surprising behavior of a curved surface interacting with a stream of air to emphasize the large gap between the state-of-the-art of experimental progress in fluid dynamics and the available theoretical account of certain phenomena. He particularly points to the inability of the theory of the time to account for those "complicated matters" (fluid dynamic problems), particularly referring to the open problem of understanding aerodynamic lift. Let's remember that the Newtonian theory for fluid resistance, known as impact theory, remained popular until the beginning of the nineteenth century.[6] According to this theory, the momentum exchange associated with the impact of fluid particles determines the resistance on an immersed body. An estimate of the net aerodynamic force on a flat plate at an angle of incidence, $\alpha$, calculated according to this theory, gives the famous $\sin^2 \alpha$ law dependence which led to the conclusion of the impossibility of practical flight for over two hundred years.[3] Moreover, even though the D'Alembert (1717-1783) paradox[7] (stated in 1768) had been solved by the end of the nineteenth century with the introduction of internal fluid friction and the Navier-Stokes equations had been formulated, the latter were far from being solvable. Accordingly, their application to actual complex problems was out of reach. Practical engineering made use of empirical equations which did not deal with "natural philosophy" considerations on the fluid's molecular interactions, the role of friction, or the boundary conditions at the solid surface. In fact, the debate about the slip or no-slip[8] of the fluid at the solid surface was still object of debate as late as 1860.[6] The no-slip condition can fail for particular conditions of the surface,[9] which are nevertheless out of the domain of practical aerodynamics.

At the beginning of last century, flight was not the only example of complex technology which had gone ahead of the existing theoretical previsions. Modern navigation and civil engineering had developed as well in the nineteenth century without a complete theoretical account. It is historically proven that theoreticians were even looked upon with suspicion, as if their predictions could even hinder the progress of practical applications of fluid dynamics. As an example, the dismantling of the Pont des Invalides in Paris which had been designed by Claude-Louis Navier (1785-1836) in 1820 demonstrated, in the words of Saint-Venant (1797-1896), that "at that time there already was a surge of the spirit of denigration, not only of the savants, but also of science, disparaged under the name of theory opposed to practice".[6]

With regard to the lack of a complete theoretical account of lift at his time, Chanute longs for "the great physicist, who, like Galileo or Newton, should bring order out of chaos in aerodynamics, and reduce its many anomalies to the rule of harmonious law".[2] This gap between practical flight and a complete aerodynamic theory was about to be filled by L. Prandtl (1875-1953), who introduced the concept of a boundary layer and thus provided the basis of modern aerodynamics progress. The resolution of the boundary layer equations, mostly done by one of Prandtl's students, Paul Richard Heinrich Blasius (1883–1970), allowed a considerable improvement in the treatment of aerodynamic problems.[10]

Now, let's consider Young's discussion on the behavior of jets of air and how it relates to what is known today as Coandă effect.

**II. YOUNG'S PAPER ON STREAMS OF AIR**

In the part II of the letter, titled "Of the direction and Velocity of a Stream of Air", Young discusses on the "divergency" of air streams blown through narrow "blowpipes". He refers to the natural crosswise widening of a turbulent jet of air when it is issued in surrounding still air.[11] Fig.1 shows an original drawing by Young in which a jet's spreading and turbulence is put into direct relation to the pressure with which the air is blown though the pipe.[12] Moreover, Young describes the qualitatively different behavior of the jet depending on its velocity at the nozzle, a behavior that he considers "extremely difficult to explain" although "it may be made distinctly perceptible to the eye, by forcing a current of smoke very gently through a fine tube". Young continues: "when the velocity is as small as possible, the stream proceeds for many inches without any observable dilatation" which defines a laminar regime, using modern terminology (see Fig. 24, in Fig.1).

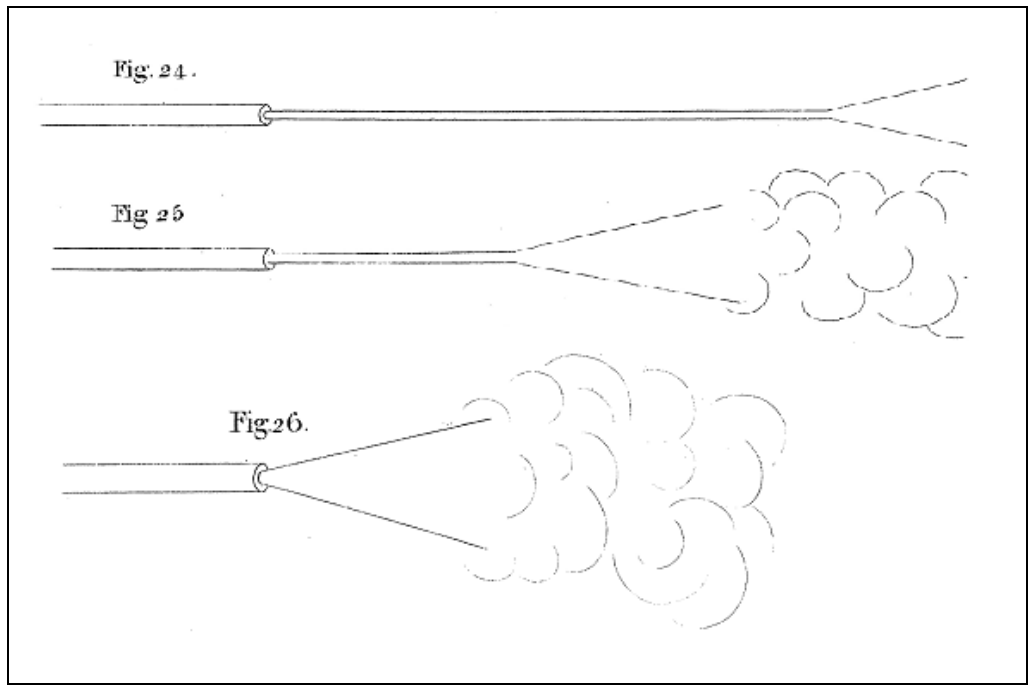

**Fig.1.** In these drawings Young shows the difference in the spreading of a jet of air and the transition to turbulence depending on the amount of pressure done on the air through the pipe. With increasing pressure (in order, from Fig. 24 to 26), the submerged jet becomes more turbulent and its spreading becomes important correspondingly closer to the nozzle.

Then, from the observation of the smoke, he continues: "it" (the air stream) "then immediately diverges at a considerable angle into a cone (…) the current seems to make something like a revolution in the surface of the cone, but this motion is too rapid to be distinctly discerned" (see Fig.25 and Fig.26, in Fig.1). The stream's divergence is manifested in an "audible and even visible vibration" according to Young's experiment. This appears to be the first description of the transition to turbulence in a flow.[13] Young anticipates, by eighty-three years, at least in the observed phenomenology, the seminal work of Osborne Reynolds[14] (1842-1912) in which he would coin for the first time the terms "*direct*" (laminar) and "*sinuous*" (turbulent) referring to water flows and the onset of the transition to turbulence. Young's further intuition is remarkable, as he wonders: "Is it not possible, that the facility with which some spiders are said to project their fine threads to a great distance, may depend upon the small degree of velocity with which they are thrown out, so that, like a minute current, meeting with little interruption from the neighbouring air, they easily continue their course for a considerable time?". Indeed, he is focusing on air's density, viscosity and speed as well as on the characteristic dimensions of the spider's thread (its diameter), or better, in the combination of these parameters, as the relevant element to understand a fluid dynamical situation. The interaction of a spider web thread with air is dominated by viscous effects and hence characterized by a low Reynolds number, as we would say in modern terms. The Reynolds number is a dimensionless parameter which describes the ratio between inertial and viscous effects in a flow. Mathematically, the Reynolds number is given by $\mathrm{Re} = \frac{\rho U L}{\mu}$ where $\rho$ and $\mu$ are, respectively, the density and dynamic viscosity of the fluid, $L$ is the characteristic dimension of the solid with which the fluid interacts (or where the fluid flows) and $U$ is the speed of the flow. High

Reynolds numbers (of the order of $10^5$ and beyond for air flows) are characteristic of situations in which viscosity plays a negligible role. Aircraft subsonic flight is an example, provided we consider the region outside the boundary layer where viscous effects can never be neglected. A back of the envelope calculation for a Boeing 787 with a wing span of 60 m (*L*) and a cruise flight speed of the order of $10^3$ km/h (*U*) gives a $\text{Re} \approx 10^8$. A spider web of characteristic length (*L*) of 7-8 *µ*m which is thread in the air at an estimated speed of the order of $10^{-3}$ m/s gives a $\text{Re} \approx 10^{-3}$. Both the airliner and the spider web interact with the same fluid but their corresponding aerodynamic regimes are completely different. The aerodynamics of an airliner and the threading of a spider web are dominated, respectively, by inertial and viscous effects.

The Reynolds number is of paramount importance in the understanding of many fluid dynamical problems, particularly regarding turbulence. This number plays a fundamental role in the working mechanism of a wind tunnel and its ability to predict the aerodynamic behavior of surfaces: the fluid dynamic behavior of two geometrically similar solids will be the same if the corresponding Reynolds numbers are of the same order of magnitude, even if the solids interact with a different fluid and have different dimensions. The use that the Wright brothers made of their studies in the wind tunnel that they had expressly constructed had great significance in their success with the Flyer in 1903.

The fact that Young is implicitly, although not mathematically, stating the role of the Reynolds number in his query about the spider web behavior, demonstrates his powerful insight. He was notoriously famous for not using mathematical equations in the description of physical phenomena, making particularly cumbersome his explanations of complex fluid dynamical phenomena which could otherwise be described with simple algebra, as stated in Craik's recent paper onYoung's contributions to fluid mechanics.[15]

To summarize, in the part II of his letter, presented to the Royal Society in 1800, Young observes several important phenomena related to the behavior of jets of air: the transition to turbulence, the importance of the combination of the relevant physical parameters (Re) to describe a fluid dynamical regime, the spreading of jets of air as a function of pressure and the role of air's inertia. The spreading of a jet of air is indeed related to a typically inertial effect, namely, the entrainment. This term refers to the phenomenon by which a submerged jet of fluid pulls along the surrounding fluid by momentum exchange. To support the explanation of this phenomenon Young discusses the experiment in which a candle flame is pulled towards a nearby stream of air (see Fig.2). According to Young, the flame's behavior is due to the "communication" of jet's momentum to air, in the direction at right angles to the jet, by "friction" (viscosity): "the flame (...) is every where forced by the ambient air towards the current, to supply the place of that which it has carried away by its friction", which is a clear explanation of the physical mechanism of entrainment.

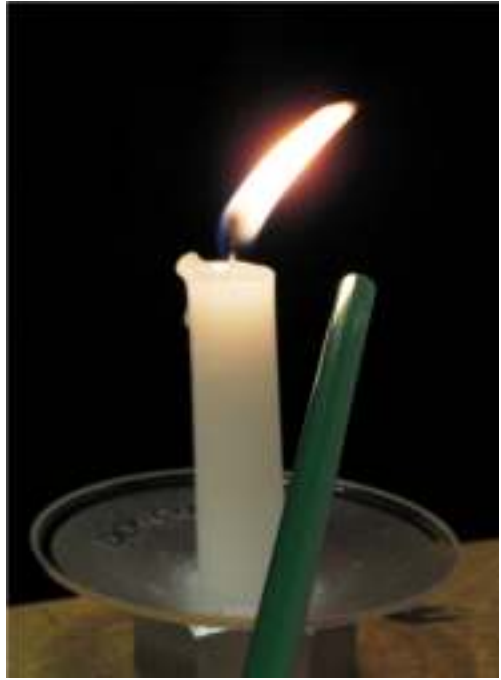

**Fig.2** A candle flame is pulled towards a thin jet of air as this entrains surrounding still air which "supplies the place of that which it has carried away by its friction" according to Young

Momentum is transferred transversally to the flow by the shear stress among contiguous layers of fluid. This can be demonstrated mathematically with a microscopic calculation obtaining the linear relation between the applied shear stress and the rate of strain, typical of Newtonian fluids.[16] Young continues: "no doubt" (the effect on the candle) "arises from the relative situation of the particles of the fluid, in the line of the current, to that of the particles in the contiguous strata, which is such as naturally to lead to a communication of motion nearly in a parallel direction; and this may properly be termed friction". Hence, he invokes the fluid's viscosity and the inertial effects generated by momentum exchange between the jet and the surrounding air to explain the observation on the candle. Young describes another interesting way to visualize the effect of a stream of air on surrounding still air, observing the ripples it generates in a surface of water: "mark the dimple which a slender stream of air makes on the surface of the water; bring a convex body into contact with the side of the stream, and the place of the dimple will immediately show that the current is inflected towards the body; and, if the body be at liberty to move in every direction, it will be urged towards the current". The jet of air is deviated by the presence of the curved surface as the heightening of the ripples demonstrates. As Chanute explains in his book on flying machines, Young demonstrated that "a certain curved surface suspended by a thread approached an impinging air current instead of receding from it" through a clear action-reaction mechanism. The force acting on the fluid through viscosity (action) generates a force on the curved surface (reaction). This experiment can be easily done using a spring scale, a party balloon and a fan. The force exerted on the balloon by the impinging jet is accordingly measured.[17] Another simple experiment showing the reciprocal effect of a jet of air acting on a curved surface, similar to that suggested by Young, is shown in Fig.3. The deviation of the jet of air is put forward with a thin plastic thread. Then, in his letter Young is describing precisely the Coandă effect and he explains it invoking the viscous behavior of air and the third principle of dynamics.

In 1910, Henri Coandă (1886-1972), a Rumanian aeronautical engineer, was testing the prototype of what would historically become the first jet airplane. By serendipity, he discovered the natural adhesion of air jets to curved surfaces, which cost him a crash during the test. In fact, the incident was due to the adhesion of the engine exhaustion gases to the fuselage on which Coandă had installed curved deflectors, hoping to achieve the opposite effect.

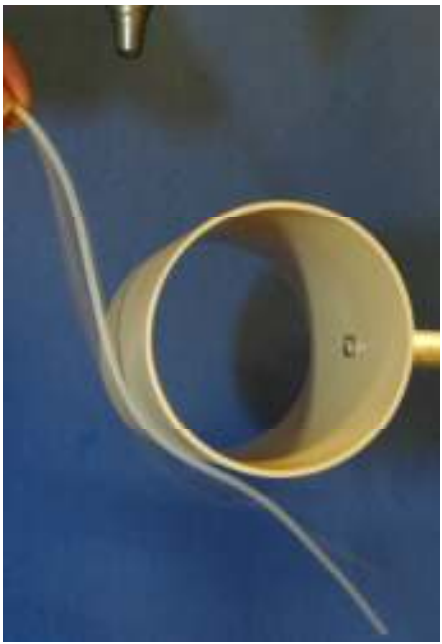

**Fig.3** A thin jet of air blown close to a curved surface attaches to it and deviates, as the plastic thread demonstrates. This kind of jet adhesion to a curved surface is known in the literature as Coandă effect

According to I. Reba,[18] Theodore Von Kármán (1881-1963) named the phenomenon after discussing it with Coandă right after his crash with the Coandă-1910. However, reading Young's letter it appears that the phenomenon had already been observed and its explanation had been given. Somehow it went unnoticed.

It is curious that another very well known fluid dynamical effect, Magnus effect,[19] which regards the aerodynamic behavior of spinning curved surfaces (a ball, a cylinder or a gun projectile) interacting with an extended flow (not a jet like in the Coandă effect) was already known at Euler's time.[3] Benjamin Robins (1707-1751) had noticed that a ballistic projectile would acquire a curved trajectory, and he attributed this effect to its spinning. However, this idea was disregarded by Euler, who believed this effect was only due to surface irregularities of the projectile. A century later Gustav Magnus (1802-1870) would "rediscover" the effect and so it bears his name today.

Most probably Von Kármán and Coandă were unaware of Chanute's and Young's references, and Von Kármán generously and righteously named the effect after the Rumanian engineer. On the other hand, Euler unintentionally deprived Robins of a well deserved notoriety, at least for the Magnus - could have been Robins – effect, particularly appreciated by soccer fans.[20,21] Nevertheless, Robins achieved an important place in the history of aerodynamics for other important contributions, not the least on transonic flow, with his experiments on ballistics.[3]

# III. PUTTING TOGETHER SOME IMPORTANT NAMES IN THE HISTORY OF AERODYNAMICS

Beyond the historical anecdote, the fact is that sometimes attributions of ideas to scientists are incorrect when the history of science is involved, especially in textbooks, as Bohren delightfully illustrates.[22] This fact regards not only the authorship of a given discovery but also the association of an explanation of a phenomenon to a given physical principle. This is particularly the case for the Coandă effect. This is sometimes improperly explained in a didactic context using Bernoulli's theorem, which cannot explain it at all, as several authors have noted.[17,23,24,25] The right explanation was already there in 1800, in the hands of one of the most famous polymaths in the history of science.[26] Young's explanation is quite simple and, most importantly, correct: a jet of air impinging on a curved surface deviates by the shear stresses acting among contiguous layers of fluid when this adheres to the surface by viscosity (non-slip condition).[27] In return, the object is pulled towards the jet: the force acting on the fluid generates a reaction in the solid. This is simple Newtonian reasoning.

The Coandă effect is an excellent paradigm, particularly in a classroom setup, to demonstrate by simple means (a fan, a light ball and a spring scale) that the deviation (curvature) of a flow due to its interaction with a solid surface is related to lift forces (intended as forces perpendicular to the flow direction) acting on the solid. However, returning to flight and the origin of lift, this paradigm should not be extended to the point of stating that lift in an airplane is generated "by Coandă effect". A wing diverts air downwards and this deviation, i.e., the asymmetry in the flow generated by the wing, is the fundamental cause of the origin of lift.[28] However, an aircraft flies in an extended flow, not in a narrow jet respect to the plane's dimensions. Jets and extended flows interact differently with solid surfaces. Accordingly, the physical explanation of Magnus and Coandă effects is fundamentally different. Both effects regard lift on a solid. However, in the Coandă effect, lift comes from a local deviation of a narrow jet of air (respect to the dimensions of the solid) while, in the Magnus effect, lift is a consequence of a fully circulatory flow around the solid, induced by its rotation. Incidentally, the Magnus effect is closer to airplane's lift than it may appear at a first glance. An airplane wing does not rotate to generate lift but, as a rotating solid, it induces air circulation around it. This circulation is related to wing's lift by the Kutta-Zhoukowsky theorem, after M. Wilhem Kutta (1867-1944) and Nikolai E. Zhukowsky (1847-1921) who first developed the mathematical theory of lift for a 2-D wing.[29]

It should not go unnoticed that Von Kármán himself, who knew and named Coandă effect, does not make any mention of it in his appealing book on the history of aerodynamics, written in 1954.[29] Regardless of its importance and countless applications,[18] this effect has nothing to do with the physical origin of lift on an airplane, as explained above. Nevertheless, it can be applied in wings, for instance, to improve boundary layer attachment and wing's performance and to avoid stalling in some circumstances.[30] All the same, it can be used to reattach flows that suffer from separation when flowing over surfaces that, contrary to aerodynamic profiles, generate adverse pressure gradients like, for instance, steps, constrictions or openings.

## IV. CONCLUSIONS

The history of the discovery of the Coandă effect is a good example of the sometimes serendipitous way in which scientific discoveries may be named or attributed to a scientist. Sometimes a phenomenon may have been already discovered by someone else, as the paper on streams of air, written in 1800 by Thomas Young, demonstrates. A paper that, in the words of its author, was intended to be a discussion on sound phenomena becomes an amazing proof of the physical insight for which Thomas Young has become famous. Young provides an accurate description of several important phenomena related to aerodynamics, lift and the onset of turbulence which continues to be a fundamental field of research nowadays.

A simple experiment, like observing the effect of a jet impinging on a ball, provides enough material to discuss the fluid dynamical differences between submerged jets and extended flows, the physical origin of lift, the role of turbulence and entrainment, and a mathematical theory of lift. One should never underestimate the importance of giving more weight in the classroom to the historical context in which scientific discoveries are made. The historical framing may help the student to appreciate the thrill and complexity of a scientific discovery, understand the human factor behind it, and make interesting connections which may help in the learning process.

a) Electronic mail: teresa@science.unitn.it

---

[1] Orville Wright, *How we invented the airplane, An Illustrated History* (David McKay Company, Inc., New York 1953). The author has read the unabridged reprint of the first edition by Dover Publications, Inc., New York, 1988

[2] Octave Chanute, *Progress in Flying Machines,* (The American Engineer & Railroad Journal, New York, 1894). The author has read the unabridged reprint of the first edition by Dover Publication, Inc, NY, 1997

[3] John D. Anderson, Jr., *A History of Aerodynamics,* (Cambridge Univ. Press, 1997)

[4] Thomas Young, "Outlines of Experiments and Inquiries Respecting Sound and Light", Phil. Trans. R. Soc. Lond. **90**, 106-150 (1800)

[5] R. Wille R and H. Fernholz H, "Report on the first European Mechanics Colloquium on the Coandă Effect ", J. Fluid Mech. **23** 801-819 (1965);

J. Feng J and D.D. Joseph, "The motion of a solid sphere suspended by a Newtonian or viscoelastic jet", J. Fluid Mech. **315** 367-385 (1996)

[6] Olivier Darrigol, *Words of Flow,* (Oxford Univ. Press, 2005)

[7] The D'Alembert paradox refers to the fact that the resistance of a body moving steadily in an inviscid fluid is zero. This is in obvious contradiction with the experience. Actually, the paradox does not subsist. The problem

arises from not considering the internal friction in the fluid due to its viscosity. However, this paradox remained an open problem well beyond the eighteenth century, until the complete formulation of the Navier-Stokes equations

[8] The absence of slip of a fluid in contact with a surface was already considered a valid hypothesis by Newton who called it "defectus lubricitatis", lack of slipperiness. Today this assumption is called no-slip condition and it is considered valid in the whole domain of classical fluid dynamics

[9] Steve Granick, Yingxi Zhu and Hyunjung Lee, "Slippery questions about complex fluids flowing past solids", Nature Materials 2, 221-227 (2003)

[10] J. D. Anderson Jr.,"Ludwig's Prandtl Boundary Layer", Phys. Today December, 42-48 (2005)

[11] Herrmann Schlichting, Klaus Gersten, *Boundary-Layer Theory*, (Springer; 8th Ed., 2000), Chapter 22

[12] This drawing belongs to the paper quoted in [4] and it is published here by kind permission of the Royal Society of London. Although reference [4] contains more drawings originally made by Young, this is the only one that it is directly related to the subject of this paper

[13] A. D.D. Craik, "Thomas Young on fluid mechanics", J Eng Math **67**, 95-113 (2010)

[14] Osborne Reynolds, "An experimental investigation of the circumstances which determine whether the motion of water shall be direct or sinuous, and of

*See, Cured the Sick, and Deciphered the Rosetta Stone, Among Other Feats of Genius,* (Pi Press, New York 2006)

[27] It should be emphasized that it is not necessary to use a curved surface in order to achieve flow curvature, provided the solid surface is positioned against the flow with a given angle of attack. This is particularly important when discussing how wings generate lift: flat and symmetrical wings can generate lift just as cambered wings do, provided they face the wind with a given angle of attack. In fact, lift in a wing is related to the flow curvature, and to the corresponding asymmetry in the flow above and under the wing, not to its camber.

[28] David F. Anderson and Scott Eberhardt *Understanding Flight* 2nd Ed. (Mc Graw Hill 2010)

[29] Theodore Von Kármán, *Aerodynamics: Selected Topics in the Light of Their Historical Development*, (Dover Pub., Inc, NY, 2004) pp 36-46

[30] G. Gerhab and C. Eastlake "Boundary layer control on airfoils" Phys. Teach. **29,** 150-151 (1991)